# Clarifying the case for retired galaxies

## Grażyna Stasińska

LUTH, Observatoire de Paris, CNRS, Université Paris Diderot; Place Jules Janssen 92190 Meudon, France (e-mail:grazyna.stasinska@obspm.fr)

**Abstract**

This is a short answer to the paper "A mid infrared study of low-luminosity AGNs with WISE" by R. Coziol et al. ( ArXiv:1405.4159v1)

**Keywords:**

galaxies: evolution -- galaxies: statistics -- galaxies: stellar content -- galaxies : active.

## 1. Introduction

In a long introduction to their paper "A mid infrared study of low-luminosity AGNs with WISE" Coziol et al. (2014, ArXiv:1405.4159v1) argue against the existence of "retired galaxies"—a denomination proposed by Stasińska et al. (2008) to describe galaxies that stopped forming stars and are ionized by "hot post-AGB stars" (PAGB). According to Stasińska et al. (2008), such galaxies can be mistaken for low luminosity AGNs (LLAGNs) in standard emission-line ratio diagrams.

To make this reply short, we will not examine in detail each of the arguments presented by Coziol et al.— most of which can be shown to be wrong—but we will better explicit what we call "retired galaxies", for the benefit of outsider readers.

In a second part, we will argue that the test proposed by Coziol et al. to refute the "PAGB hypothesis" as an explanation of objects identified as LLAGNs is incorrect.

## 2. PAGB, hot white dwarfs and HOLMES

It seems that part of the misunderstanding comes for the terminology we used. Coziol et al. stick to the concepts of post-AGB stars, planetary nebulae and white dwarfs that are in usage in various fields of astronomy when identifying and studying these objects. For example, in Binette et al. (1994), which was the first paper proposing the existence of what was later called "retired galaxies", we used the terms "old stars" and "post-AGB" stars. Post-AGB was obviously taken in a broad meaning, similar to the one used by Renzini (1985), Greggio & Renzini (1990), or Bloecker (1995) for example. In the meantime, the term of post-AGB became increasingly popular to refer to the stellar evolution phase just after the asymptotic giant branch and before the planetary nebula phase, as in van Winckel (2003) or Szczerba et al. (2007). This is why in Stasińska et al. (2008) we wrote that "the ionization in these "retired" galaxies would be produced by hot post-asymptotic giant branch stars and white dwarfs". This was not a fortunate choice of terminology, being too specific. This is why in Cid Fernandes et al. (2011, see also Flores-Fajardo et al 2011) we opted for the use of the terminology "hot evolved low-mass stars (HOLMES)". This includes possible AGB-manqués (Ferguson & Davidsen 1993) which may enter the ionizing photons budget, but excludes all kinds of massive stars.

That old stellar populations produce ionizing photons is undebatable considering the present-day knowledge of stellar evolution theory. What can be debated, though, is the total luminosity of ionizing photons produced. Stellar evolution after the asymptotic giant branch is a notoriously difficult subject, and computations of stellar evolution tracks require approximations. In Binette et al. (1994) we used the Bruzual & Charlot (1993) stellar population synthesis tracks, in Stasińska et al. (2008) we used the Bruzual & Charlot (2003) ones, and in Cid Fernandes et al. (2011) we made a direct comparison with tracks from other groups (Fioc & Rocca-Volmerange 1997, Molla et al. 2009). All of them roughly agree in their prediction of ionizing photon fluxes.

What we argue is that, with our present state of knowledge, the old stellar population in galaxies that have stopped forming stars are providing a sufficiently large number of ionizing photons to explain the observed Hα luminosities and equivalent widths in many galaxies previously mistaken for low luminosity AGNs.

This does not mean that low luminosity AGNs do not exist. Simply, while in some galaxies they overtake the ionization by old stellar populations, in others they may be completely dominated by them. The frontier between weak AGNs and retired galaxies proposed by Cid Fernandes et al. (2011) using their WHAN diagram is purely empirical and only of statistical value. With more detailed studies on large and complete samples of galaxies, such as spectroscopy with integral field units or X-ray observations, this frontier could perhaps be better defined.

Incidentally note that, in their use of the WHAN diagram, Coziol et al. erroneously draw the separation between LINERs and retired galaxies at an Hα equivalent width of 0.5Å, while Cid Fernandes et al. (2011) defined it to be at 3Å. This explains why Coziol et al. write: "there seem to be very few LLAGNs in our sample consistent with the PAGB hypothesis."

In any case, it is not possible to refute the existence of retired galaxies using the arguments presented by Coziol et al. (2014). The *entire* old stellar population must be modelled, and considerations on only one of their components is misleading. Let us note, for example, that contrary to their claim that "one needs then to establish what fraction of the ionizing photons are escaping the nebulae", this is not necessary as far as the total Hα emission is concerned: whether the ionizing photons are absorbed in a planetary nebula or in the diffuse interstellar medium does not change the total Hα emission---as long as there is sufficient gas present in the galaxy to absorb them.[1]

## 3. Coziol's diagnostic diagram in the infrared

Coziol et al. (2014) write: "In the previous section we have shown that the MIR colors of the LLAGNs are significantly different from those of the PAGBs and PNs. Only a few of these stars have colors consistent with those of the LINERs and LLAGNs in our sample. However, for the PAGB hypothesis to be satisfied, the PAGBs and PNs must define the colors of the LLAGNs, which implies that their color distributions must be exactly the same. […] We conclude that the PAGB hypothesis cannot be maintained on the basis of the data we presented".

Galaxies are aggregates of stars in all evolutionary stages, and their colours are given by the *sums* of fluxes in the different bands from the individual stars. It makes no sense to compare them to colours of *individual* stars.

Note that the MIR colours of galaxies which we classify as retired are similar to colours of elliptical galaxies in general. And MIR colours of galaxies which we classified as weak AGNs (Cid Fernandes et al. 2011) are intermediate between those of strong AGNs and retired galaxies.

---

[1] As a matter of fact, there are galaxies with very similar properties to retired galaxies that, for reasons yet to be elucidated, show no emission lines at all.